\documentclass[twocolumn,,preprintnumbers,amsmath,amssymb, superscriptaddress, prb]{revtex4-1}

\usepackage{multirow}
\usepackage{graphicx}
\usepackage{textcomp}
\usepackage{color}
\usepackage{float}
\usepackage{eqnarray,amsmath}
\definecolor{red}{rgb}{1,0,0}
 
\definecolor{green}{rgb}{0,1,0}
 
\definecolor{blue}{rgb}{0,0,1}

\newcommand{\beq}{\begin{equation}}
\newcommand{\eneq}{\end{equation}} 
\newcommand{\beqa}{\begin{eqnarray}}
\newcommand{\eneqa}{\end{eqnarray}} 
 
\newcommand{\bta}{\begin{tabular}}
\newcommand{\enta}{\end{tabular}}

\begin{document}

\title{Multiferroicity in the geometrically frustrated FeTe$_2$O$_5$Cl}

\author{M. Pregelj}
\affiliation{Jo\v{z}ef Stefan Institute, Jamova c.\ 39, SI-1000 Ljubljana, Slovenia}

\author{A. Zorko}
\affiliation{Jo\v{z}ef Stefan Institute, Jamova c.\ 39, SI-1000 Ljubljana, Slovenia}
\affiliation{EN--FIST Centre of Excellence, Dunajska 156, SI-1000 Ljubljana, Slovenia}

\author{O. Zaharko}
\affiliation{Laboratory for Neutron Scattering, PSI, CH-5232 Villigen, Switzerland}

\author{P. Jegli\v{c}}
\affiliation{Jo\v{z}ef Stefan Institute, Jamova c.\ 39, SI-1000 Ljubljana, Slovenia}
\affiliation{EN--FIST Centre of Excellence, Dunajska 156, SI-1000 Ljubljana, Slovenia}

\author{Z. Kutnjak}
\affiliation{Jo\v{z}ef Stefan Institute, Jamova c.\ 39, SI-1000 Ljubljana, Slovenia}

\author{Z. Jagli\v{c}i\'{c}}
\affiliation{Institute of Mathematics, Physics and Mechanics, Jadranska c.\ 19, SI-1000 Ljubljana, Slovenia}

\author{S. Jazbec}
\affiliation{Jo\v{z}ef Stefan Institute, Jamova c.\ 39, SI-1000 Ljubljana, Slovenia}

\author{H. Luetkens}
\affiliation{Laboratory for Muon-Spin Spectroscopy, PSI, CH-5232 Villigen, Switzerland}

\author{A. D. Hillier}
\affiliation{ISIS Facility, Rutherford Appleton Laboratory, Chilton, Didcot, Oxon OX11 OQX, United Kingdom}

\author{H. Berger}
\affiliation{Institut de Physique de la Mati\`{e}re Complexe, \'{E}cole Polytechnique F\'{e}d\'{e}rale de Lausanne, CH-1015 Lausanne, Switzerland}

\author{D. Ar\v{c}on}
\affiliation{Jo\v{z}ef Stefan Institute, Jamova c.\ 39, SI-1000 Ljubljana, Slovenia}
\affiliation{Faculty of mathematics and physics, University of Ljubljana, Jadranska c.\ 19, SI-1000 Ljubljana, Slovenia}

\date{\today}

\begin{abstract}

The layered FeTe$_2$O$_5$Cl compound was studied by specific-heat, muon spin relaxation, nuclear magnetic resonance,  dielectric, as well as neutron and synchrotron x-ray diffraction measurements, and the results were compared to isostructural FeTe$_2$O$_5$Br.
We find that the low-temperature ordered state, similarly as in  FeTe$_2$O$_5$Br, is multiferroic -- the elliptical amplitude-modulated magnetic cycloid and the electric polarization simultaneously develop below 11\,K.
However, compared to FeTe$_2$O$_5$Br, the magnetic elliptical envelop rotates by 75(4)\,$^\circ$ and the orientation of the electric polarization is much more sensitive to the applied electric field.
We propose that the observed differences between the two isostructural compounds arise from geometric frustration, which enhances the effects of otherwise subtle Fe$^{3+}$ ($S$\,=\,5/2) magnetic anisotropies.
Finally, x-ray diffraction results imply that, on the microscopic scale, the magnetoelectric coupling is driven by shifts of the O$_1$ atoms, as a response to the polarization of the Te$^{4+}$ lone-pair electrons involved in the Fe-O-Te-O-Fe exchange bridges.

\end{abstract}


\maketitle

\section{Introduction}

Geometrically frustrated magnetoelectric (ME) multiferroics, where complex magnetic long-range order (LRO) induces the electric polarization, may exhibit spectacularly strong ME coupling.\cite{Cheong, Khomski, Eerenstein, Wang09}
The applied magnetic field can influence the magnetic LRO and can thus through ME coupling change the electric polarization. 
Vice versa, the magnetic order may be manipulated by external electric field, acting on the electric polarization.\cite{Cheong}
Magnetic frustration is necessary for the realization of complex magnetic structures\cite{Mendels} without the inversion symmetry, which allow for the establishment of the electric polarization.\cite{Wang09}
Since the geometric frustration arises from the arrangement of the magnetic ions within the crystal lattice,\cite{Mendels} the understanding of the relation between structural and magnetic properties is vital for the design of materials with an enhanced ME effect.\cite{Wang09}

An established approach, which has largely contributed to the knowledge about the ME phenomenon, is based on the investigation of the isostructural relatives of already known multiferroics.\cite{Goto2004,Hur2004,Kimura2007,Hollmann2010,Drechsler2007} 
Isostructural compounds may exhibit very different ME responses, depending on the leading lattice parameter involved in the ME mechanism.
The latter has two main origins; (i) changes of the angle between atoms in the superexchange bridge that affects the strength of the corresponding interactions,\cite{Goto2004} and (ii) changes of the local structure in the vicinity of the magnetic ion, determining its local crystal-field anisotropy.\cite{Hur2004,Kimura2007,Hollmann2010}
As a result, significantly different magnetic states can be realized, from which only a few may be multiferroic.
The prominent examples of such experimental approach are studies of $R$MnO$_3$, where the radius of the rare-earth ion $R$ influences the Mn-O-Mn angle and thus selects between non-ferroelectric collinear and multiferroic spiral magnetic ground states.\cite{Goto2004}

\begin{figure} [!]
\includegraphics[width=0.50\textwidth]{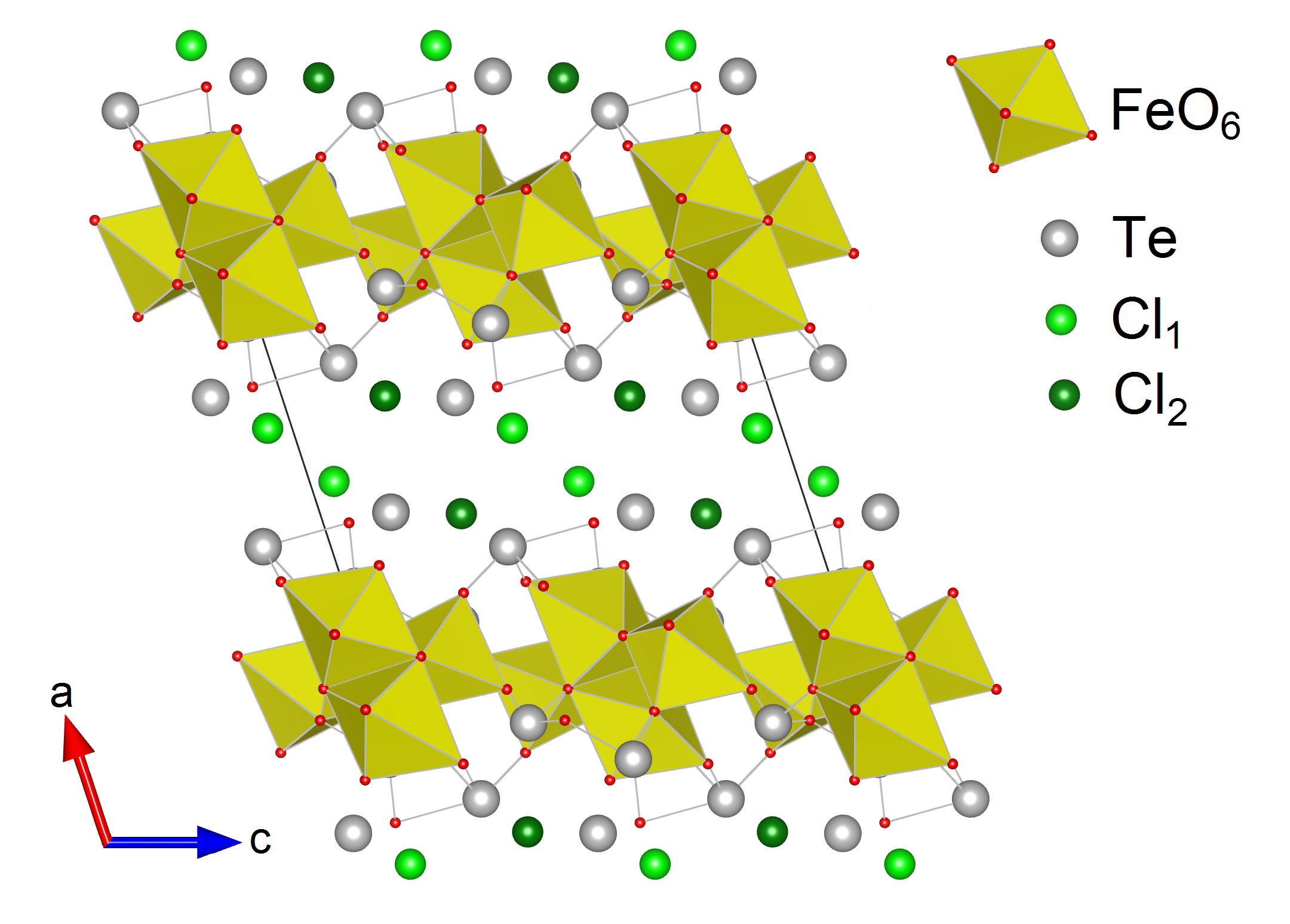}
\caption{(Color online) Crystal structure of the layered FeTe$_2$O$_5X$, $X$\,=\,Cl (or Br). The $bc$ layers are separated by Cl$^-$ ions (light green for Cl$_1$ and dark green for Cl$_2$), which determine the interlayer distance. Te$^{4+}$ ions are gray (large spheres), O$^{2-}$ are red (small spheres), while [FeO$_6$] are shown as yellow octahedra.}
\label{crystruct}
\end{figure}
Here we focus on the FeTe$_2$O$_5$Cl$_x$Br$_{1-x}$ family of layered materials with monoclinic unit cell (space group $P2_1 /c$ at room temperature), where $x$ can be varied between 0 and 1 (Fig.\,\ref{crystruct}).\cite{Becker}
The layers, oriented perpendicular to $a^*$, consist of [Fe$_4$O$_{16}$]$^{20-}$ iron tetramers connected via Te$^{4+}$ ions and are separated by halogen ions located at two crystallographically non-equivalent positions.\cite{Becker}
Up to now, the majority of studies\cite{PregeljPRL09,PregeljPD,PregeljPRL12,PregeljNMR} - including dielectric and magnetic structure investigations - have been restricted to $x$\,=\,0, namely FeTe$_2$O$_5$Br (FTOB), which was found to be multiferroic at low temperatures.\cite{PregeljPRL09}

The density-functional-theory calculations yield a surprising result that the intra-tetramer Fe-O-Fe and the inter-tetramer Fe-O-Te-O-Fe exchange interactions are comparable.\cite{PregeljAFMR}
Consequently, the system acts as an assembly of alternating antiferromagnetic Fe$^{3+}$ ($S$\,=\,5/2) spin chains running along the $c$ axis coupled by weaker interactions, which are geometrically frustrated.\cite{PregeljAFMR}
Moreover, Te$^{4+}$ ions possess lone-pair electrons, which can be easily polarized and can thus lead to the electric polarization, if the inversion symmetry of the $P2_1 /c$ structure is broken.\cite{LonePair, Cheong}
In FTOB, two magnetic transitions occur, at $T_{N1}$\,=\,11\,K and $T_{N2}$\,=\,10.5\,K, separating the high- and the low-temperature incommensurate (HT-IC and LT-IC) magnetic LRO phases.\cite{PregeljPD} 
Both phases exhibit elliptical amplitude-modulated magnetic LRO, where a substantial part of the Fe$^{3+}$ magnetic moments is fluctuating down to the lowest experimentally accessible temperatures.\cite{PregeljPRL12}
Below $T_{N2}$, in the LT-IC phase, spontaneous electric polarization develops,\cite{PregeljPRL09}
indicating a strong ME coupling, which is evidenced also by a magnetic field dependence of $T_{N2}$.\cite{PregeljPD}
The ME coupling mechanism was explained by phase shifts between exchange coupled spin modulation waves.\cite{PregeljPRL09,PregeljNMR} 
However, since the crystal-lattice distortions accompanying the multiferroic transition are extremely small,\cite{PregeljPRL09,PregeljAFMR} complete understanding of this exotic ME mechanism on the microscopic level is not yet achieved.
For instance, it is still not clear to what extent the lone-pair electrons are involved in the ME coupling or why electric polarization develops only below $T_{N2}$ even though symmetry restrictions are broken already in the HT-IC phase.\cite{PregeljNMR}

To address these questions we focus on the isostructural FeTe$_2$O$_5$Cl (FTOC) compound ($x$\,=\,1), where the smaller size of the halogen ion (Cl$^-$ has by $\sim$0.15\,\AA\ smaller radius compared to Br$^-$) leads to substantially reduced interlayer distance.\cite{Becker} 
On the other hand, the interatomic distances within each layer are very similar to those in FTOB, implying that the dominant intralayer exchange interactions in both systems are comparable.
This agrees with the Curie-Weiss temperature $\theta$ = --124\,K, which is close to --98\,K measured in FTOB.\cite{Becker}
Moreover, the magnetic transition temperatures, $T_{N1}$\,=\,12.6\,K and $T_{N2}$\,=\,11\,K,\cite{Becker} are also very similar to the values found in FTOB.\cite{PregeljPD}
Interestingly, far-infrared study of FTOC could not detect any link between electrodynamic response and magnetic ordering at low temperatures,\cite{Pfuner} implying that, in contrast to FTOB, FTOC may not be multiferroic.

Here we present a comprehensive study of the FTOC single crystals including specific-heat, muon spin relaxation ($\mu$SR), nuclear magnetic resonance (NMR), neutron diffraction, dielectric and synchrotron x-ray diffraction measurements, allowing us a full characterization of magnetic ordering, dielectric response and the coupling between them.
Results provide important new data on the FeTe$_2$O$_5X$ family and improve the understanding of the irregular ME mechanism active in these systems.

\section{Experimental Details}

High-quality powder and single crystals of FTOC were grown by the standard chemical-vapor-phase method.\cite{Becker}

Specific-heat measurements were performed between 4 and 16\,K on a Quantum Design Physical Property Measuring System at Jo\v{z}ef Stefan Institute (JSI), Slovenia.

The $\mu$SR spectra were measured on powder samples in zero field as well as in a weak transverse field of 3\,mT between 2 and 50\,K on the General Purpose Surface-Muon instrument at the Paul Scherrer Institute (PSI), Switzerland.
Additional  zero-field experiments between 60\,mK and 5\,K were conducted on the MuSR instrument at the ISIS facility, Rutherford Appleton Laboratory, United Kingdom.

The $^{35}$Cl NMR was measured using a home-built spectrometer at JSI between 4 and 300\,K in the magnetic field of 9.4\,T applied along $a^*$, i.e., the reference Larmor frequency is $^{35}\nu_0$\,=\,39.192\,MHz.

Single-crystal neutron diffraction experiments with the wave length 2.317\,\AA\, were performed on a plate-like crystal of the size 12$\times$8$\times$2\,mm$^3$ between 2 and 20\,K on the single-crystal diffractometer TriCS at PSI.
Magnetic structure was refined from integrated intensities of the magnetic reflections using MagOpt program based on the CrysFML library.\cite{ CrysFML}

The complex dielectric constant, $\epsilon^*$\,=\,$\epsilon'(T)$\,--\,$\epsilon''(T)$, was measured as a function of temperature and frequency by using an HP4282A precision LCR meter at JSI.
The dielectric constant was scanned at a few frequencies between 20\,Hz and 1\,MHz on cooling or heating the sample with the typical cooling/heating rates of 10\,K/h in various dc bias electric fields ranging from 0 to 3\,kV/cm.
The excitation electric ac field of 100-400\,V/cm was applied along the $a^*$, $b$, and $c$ axes.
Zero-field ac dielectric measurements and ac dielectric measurements in the dc electric bias field were performed in an Oxford continuous-flow liquid-helium cryostat.

Single-crystal x-ray synchrotron diffraction data were acquired at the BM01A Swiss-Norwegian Beamline of ESRF, Grenoble, France. Data sets ($\sim$ 780 reflections per temperature point) were collected between 4.5 and 35\,K at the wavelength of 0.64\,\AA, using a closed-cycle He cryostat mounted on a six-circle kappa diffractometer (KUMA), while the interatomic distances were refined using the SHELX97 program.\cite{SHELX}

\section{Results}

\subsection{Specific-heat measurements}

\begin{figure} [!]
\includegraphics[width=0.50\textwidth]{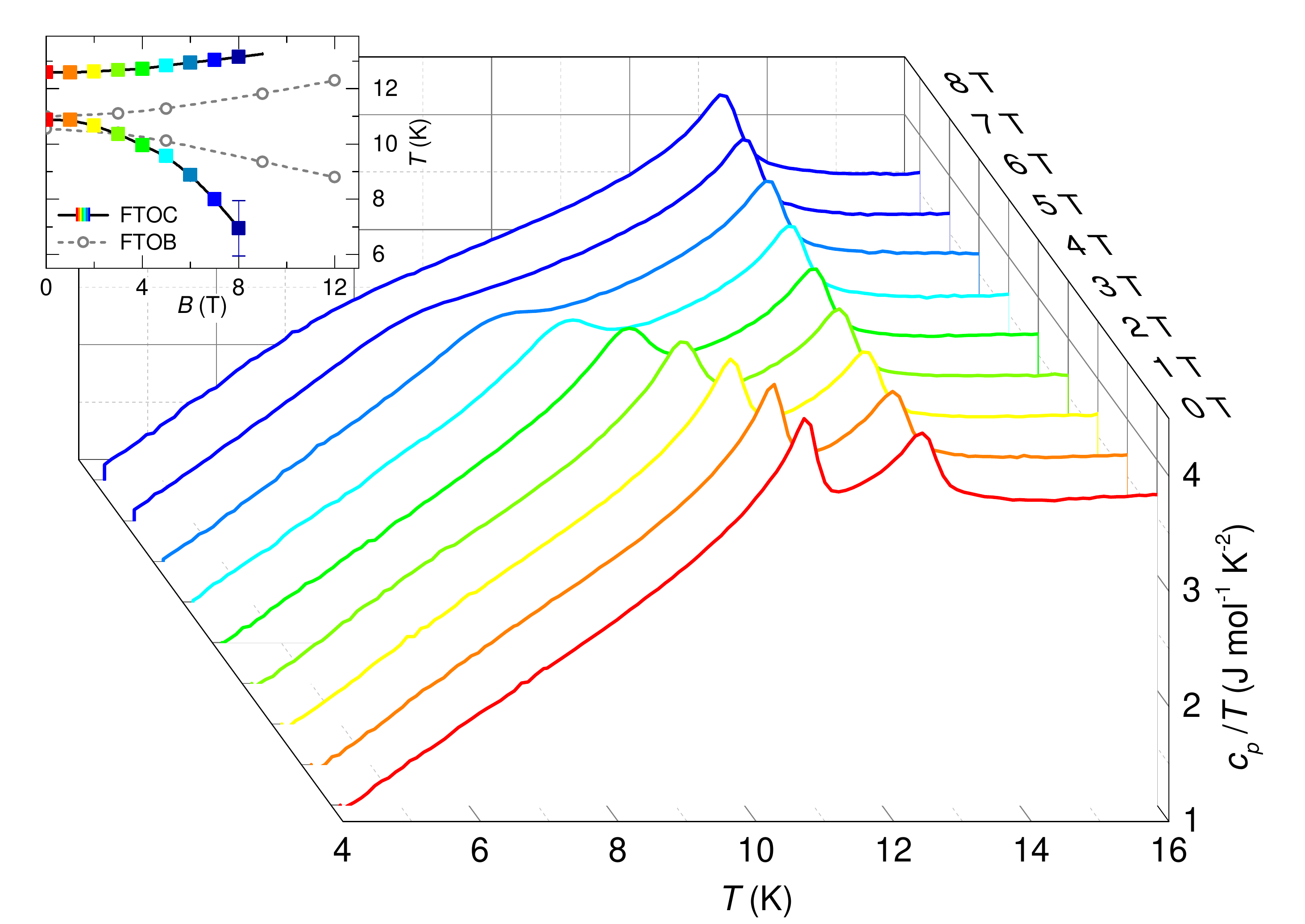}
\caption{(Color online) Temperature dependences of the specific heat divided by temperature in increasing magnetic fields along $a^*$ axis. Inset: Derived field dependence of the magnetic transitions (solid symbols -- the colors match the colors of the corresponding experimental curves) in comparison to their behavior in FTOB (open symbols). Lines are guides for the eyes.}
\label{cp}
\end{figure}

Specific-heat measurements reveal two $\lambda$-type anomalies, which in the absence of the external magnetic field ($B$\,=\,0) appear at $T_{N1}$\,=\,12.5\,K and $T_{N2}$\,=\,10.9\,K  (Fig.\,\ref{cp}), i.e., in accordance with the previous study.\cite{Becker} 
As in the isostructural FTOB system, the two anomalies reflect the corresponding transitions from the paramagnetic to the high- and to the low-temperature (HT and LT) magnetic phases.\cite{PregeljPD}
When magnetic field $B$ is applied along the $a^*$ axis, the lower anomaly, at $T_{N2}$, broadens and shifts with increasing $B$ to lower temperatures.
The observed $T_{N2}(B)$ dependence is much more pronounced than in FTOB (inset in Fig.\,\ref{cp}) and thus implies that the low-temperature magnetic ordering in FTOC is more sensitive to the applied magnetic field, making the compound more appropriate to study a potential ME response.
On the other hand, the anomaly at $T_{N1}$ gradually shifts to higher temperatures with increasing field, reaching $T_{N1}$\,=\,13.20(3)\,K at 8\,T, i.e., reminiscent of the behavior in FTOB.

\subsection{$\mu$SR measurements}

Although, $T_{N1}$ and $T_{N2}$ in FTOC show similar field dependence as in FTOB, the magnetic ordering is yet unknown.
To address this issue, we first conducted local-probe $\mu$SR experiments.
The muons are positively charged particles with a large magnetic moment, making them highly sensitive probes of local magnetic fields. 
Hence, $\mu$SR can easily distinguish between fluctuating and static magnetism as well as between LRO and static magnetic disorder.\cite{muSR-book}

\begin{figure} [!]
\includegraphics[width=0.5\textwidth]{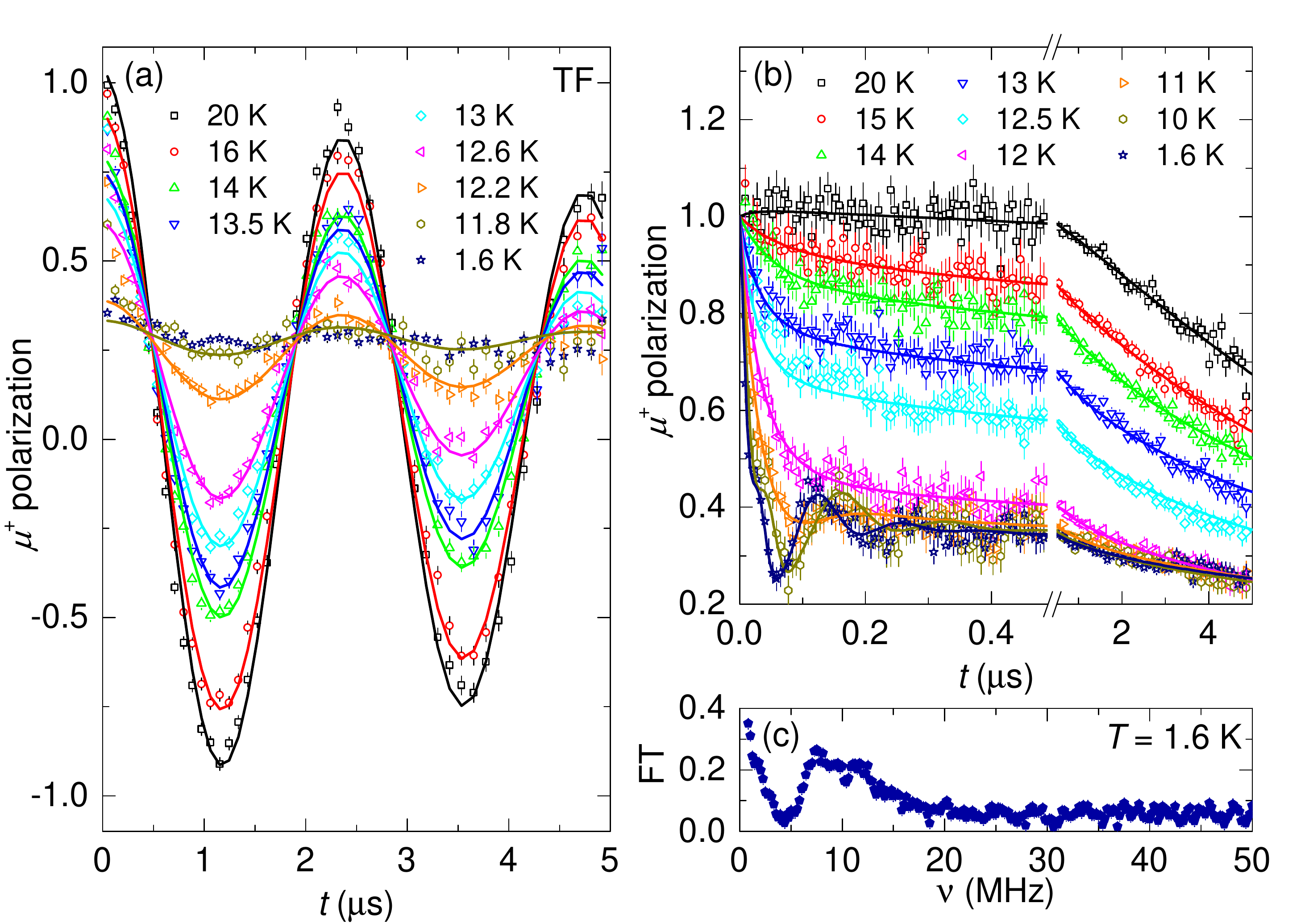}
\caption{(Color online) Temperature dependence of the muon relaxation in (a) a weak transverse field of 3\,mT, with corresponding fits (solid lines, see text for details), and (b) in zero field. Solid lines are fits to the two component model. (c) Fourier transform of the $\mu$SR signal in zero field at 1.6\,K.}
\label{muSR-spec}
\end{figure}
To trace the evolution of the static magnetic correlations, we first performed $\mu$SR measurements in a weak transverse magnetic field (TF) $B_{\text{TF}}$\,=\,3\,mT.
In the paramagnetic phase, i.e., well above $T_{N1}$, the static field $B$\,=\,$B_{\text{TF}}$ exceeds other (nuclear) static fields, so the muon precession frequency $\nu$\,=\,$\gamma_\mu B/2\pi$ is exactly determined by $\gamma_\mu$\,=\,2$\pi\times$135.5\,MHz/T, reflecting in an oscillating $\mu$SR signal  $A(t)$\,=\,$A_0\cos(\gamma_\mu B_{\text{TF}} t)$ [Fig.\,\ref{muSR-spec}(a)].
When static spin correlations start to develop, the amplitude of these oscillations is suppressed by the internal field $B_{\text{int}}$, developing at the $\mu^+$ site and largely exceeding $B_{\text{TF}}$, i.e., $B_{\text{int}}$\,$\gg$\,$B_{\text{TF}}$.
Below $T_{N1}$, this clearly reflects in the development of the so-called ''1/3 tail'', which in powder samples originates from the muons in the magnetically ordered phase whose initial polarization is parallel to the internal field.
Consequently, the TF signal for $t$\,$>$\,0.5\,$\mu$s can be modeled as\cite{muSR-book} 
\begin{equation}
A_{\text{TF}}(t)= (1-v_m)\cos(\gamma_\mu B_{\text{TF}} t)\text{e}^{-(\lambda t)^\alpha} +\frac{1}{3}v_m.
\end{equation}
Here $v_m$ denotes the volume fraction of the sample, where magnetic correlations are significant ($B_{\text{int}}$\,$\gg$\,$B_{\text{TF}}$), $\lambda$ is the corresponding relaxation rate governed by local-field fluctuations, and $\alpha$ is a stretch exponent, which is for quickly fluctuating electronic fields, e.g., in paramagnetic phase, typically close to 1, while it can be substantially reduced when relaxation rates are distributed.
\begin{figure} [!]
\includegraphics[width=0.50\textwidth]{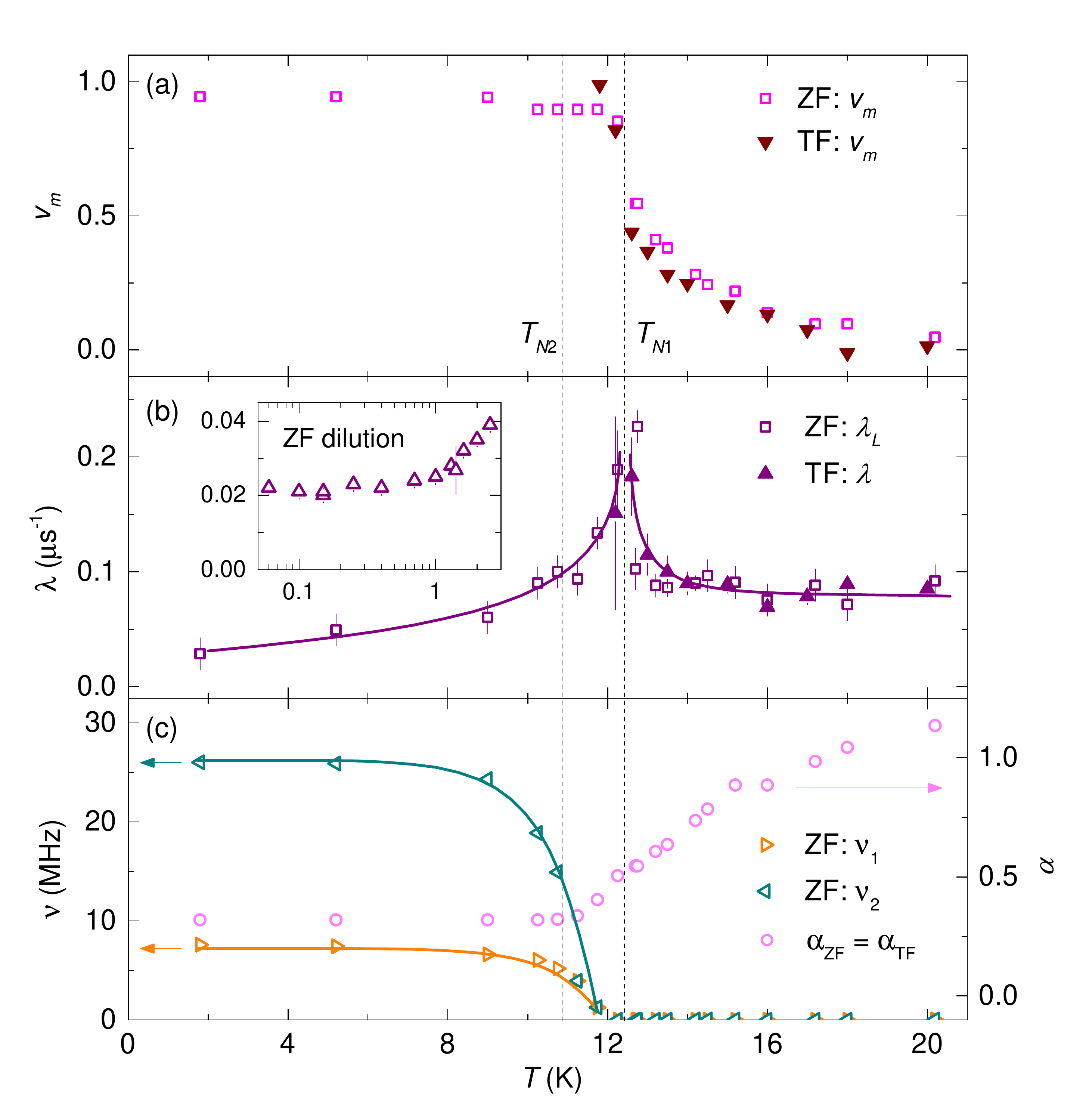}
\caption{(Color online) Temperature dependencies of (a) the magnetic volume fraction $v_m$, (b) the longitudinal relaxation rate, (c) the $\mu$SR frequencies $\nu_i$ ($i$\,=\,1,2) and the stretch exponent $\alpha$, determined from the weak transverse-field (TF) and the zero-field (ZF) data. Inset in (b) shows longitudinal relaxation for ZF experiment in a dilution refrigerator, whereas solid lines are guides for the eyes.}
\label{muSR-par}
\end{figure}
Below 20\,K, $v_m$ increases and finally reaches $\sim$1 at $T_{N1}$ [Fig.\,\ref{muSR-par}(a)].
At the same time, $\lambda$ increases in a diverging manner when approaching $T_{N1}$ [Fig.\,\ref{muSR-par}(b)], while  $\alpha$ reduces from $\sim$1 to 0.45(2) at 12\,K [Fig\,\ref{muSR-par}(c)].
The observed behavior indicates the growth of short-range correlations already above the magnetic transition and the establishment of the magnetic LRO below $T_{N1}$, where $B_{\text{int}}$ is substantially stronger than $B_{\text{TF}}$ and thus suppresses the amplitude of the corresponding oscillation of the $\mu$SR signal.

To probe the magnetic LRO we switch to the zero-field (ZF) experiment. 
On cooling below 20\,K the $\mu$SR signal exhibits a monotonic stretched exponential decay,\cite{muSR-book} in agreement with the TF results.
At $T_{N1}$, strongly damped high-frequency oscillations start to develop [Fig.\,\ref{muSR-spec}(b)], proving the presence of static internal fields of the electronic origin.
Strong damping implies a broad local-field distribution, which is clearly manifested also in the Fourier transform of the time-dependent part of the signal $A(t)$\,--\,1/3 [Fig.\,\ref{muSR-spec}(c)].
The response is reminiscent of that of isostructural FTOB with the IC magnetic ordering.\cite{PregeljPRL12}
Since the $\mu$SR study of FTOB revealed two muon stopping sites,\cite{PregeljPRL12} we likewise model our ZF $\mu$SR spectra by the two-component empirical model\cite{Zorko}
\begin{align} \label{2model}
A_{\text{ZF}}(t) =& \,\frac{2}{3}v_m\sum_{i=1}^2 \left[A_0^{(i)} \cos(\gamma_\mu B_{(i)} t)\text{e}^{-\lambda_T^{(i)} t}\right]  \nonumber \\
&+ (1-\frac{2}{3}v_m)\text{e}^{-(\lambda_L t)^\alpha}.
\end{align}
This expression accounts for the relaxation at the two muon stopping sites ($i$\,=\,1,2) with the corresponding fractions $A_0^{(i)}$ ($\sum_{i=1}^2A_0^{(i)}$\,=\,1) and the mean oscillation frequencies $\nu_i$\,=\,$\gamma_\mu B_{(i)}$ that are determined by average local magnetic fields $B_{(i)}$. 
The longitudinal relaxation rate $\lambda_L$ accounts for the spin fluctuations responsible for the stretch-exponential decay of the $\mu$SR tail, while in addition to spin fluctuations, transverse relaxation rate $\lambda_T^{(i)}$ can effectively detect also finite local-field distributions.
By fitting the complete temperature dependence of the ZF signal to Eq.\,\eqref{2model} [Fig.\,\ref{muSR-spec}(b)], we find that below 20\,K $v_m$ gradually changes from 0 to $\sim$1, where it settles at $T$\,$\leq$\,$T_{N1}$ [Fig\,\ref{muSR-par}(a)].
Similarly, the stretch exponent $\alpha$ reduces from $\sim$1 to 0.3 [Fig\,\ref{muSR-par}(c)].
This indicates a transition from the paramagnetic phase to a magnetic LRO state with a broad relaxation-rate distribution, i.e., in agreement with TF results.
Moreover, the magnetic transition is also reflected in $\lambda_L(T)$, displaying a sharp $\lambda$-type anomaly at $T_{N1}$ [Fig\,\ref{muSR-par}(b)].
We note that the temperature dependence of $\alpha$, was correlated with the fit of the TF signal, in order to ensure the comparability of the derived parameters.
The derived $\nu_i(T)$, on the other hand, displays the evolution of the magnetic order parameter in the magnetic LRO phase, as $B_{(i)}$ is directly related to the ordered magnetic moment.
We stress that $\nu_{i}$ values at low temperatures ($T$\,$\ll$\,$T_{N1}$) are in agreement with those determined for FTOB,\cite{Zorko} signifying further similarities between the two magnetic orders.
Finally, this correspondence reflects in very fast relaxation rates $\lambda_T^{(1)}$ and $\lambda_T^{(2)}$ that stretch between 15 and 60\,$\mu$s$^{-1}$, accounting for the broad local-field distribution [Fig.\,\ref{muSR-spec}(c)], which was found also in FTOB.\cite{PregeljPRL12}

The fact that $\lambda_L$\,$\neq$\,0 endures even at 1.6\,K, indicating that spin fluctuations are still active deep in the magnetically ordered state, prompted us to further cool the sample.
We performed additional ZF $\mu$SR measurements down to 60\,mK [Inset in Fig.\,\ref{muSR-par}(b)]. 
Since $\alpha$\,=\,0.3 was found to be constant below 8\,K [Fig.\,\ref{muSR-par}(c)], we kept this value fixed in the analysis in the whole low-temperature region between 2\,K and 60\,mK.
As in FTOB, $\lambda_L(T)$ saturates below 1\,K and settles at a finite value of $\sim$0.020(3)\,$\mu$s$^{-1}$, implying the existence of the persistent spin dynamics - a generic property of IC amplitude modulated magnetic order.\cite{PregeljPRL12}
The limiting zero-temperature relaxation rate is notably higher than in FTOB, where it is 0.010(5)\,$\mu$s$^{-1}$, suggesting that the density of gapless magnetic excitations is in FTOC higher.

\subsection{NMR measurements}

\begin{figure} [!]
\includegraphics[width=0.50\textwidth]{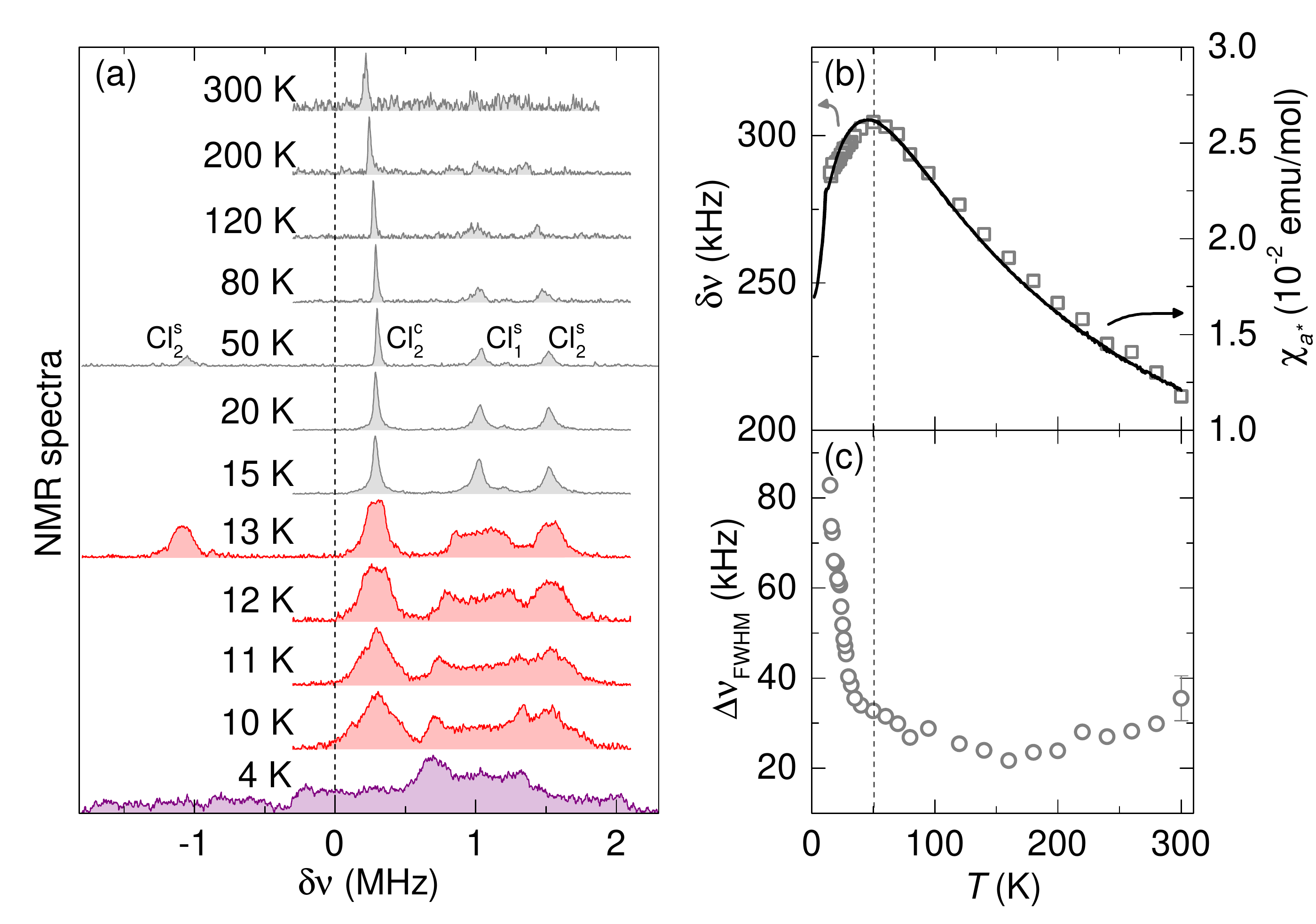}
\caption{(Color online) (a) Temperature evolution of the $^{35}$Cl NMR spectrum in the magnetic field of 9.4\,T, applied along $a^*$. The superscripts s and c denote satellite and central Cl$_i$ ($i$\,=\,1,2) NMR transitions, respectively, while different colors are used for different magnetic phases. Temperature dependences of (b) the position and (c) the linewidth of the central Cl$_2$ paramagnetic signal. The dotted line denotes the temperature where $\delta\nu$ has a maximum. Solid line in (b) is the magnetic susceptibility measured in field of 0.01\,T applied along $a^*$.}
\label{NMR}
\end{figure}

NMR is another highly sensitive local-probe technique that probes magnetic properties.
In addition, when the investigated nucleus has a spin $I$\,$>$\,1/2, it is characterized by a finite quadrupole moment (Q), through which it couples to electronic-field gradient (EFG), making it sensitive to the tiniest changes of the crystal structure.
To simultaneously probe both, magnetic as well as structural (potentially dielectric) responses, we performed NMR measurements on the $^{35}$Cl nuclei ($I$\,=\,3/2) in the magnetic field of 9.4\,T applied along $a^*$ in the frequency range $\pm2.5$\,MHz around the $^{35}$Cl Larmor frequency, $^{35}\nu_0$. 
The Cl ions are located in-between the crystal layers at two crystallographically non-equivalent sites, namely Cl$_1$ and Cl$_2$ (Fig.\,\ref{crystruct}), which have no special site symmetries and where their EFGs should be significantly different, as found for Br$_1$ and Br$_2$ in isostructural FTOB.\cite{PregeljNMR} 
Finally, we note that the $^{37}$Cl isotope has $\sim$5 times smaller absolute sensitivity and almost 20\,\% lower resonant frequency (by 6.57\,MHz at 9.4\,T) than $^{35}$Cl, hence all signals shown in Fig.\,\ref{NMR} are ascribed to the $^{35}$Cl isotope.

At room temperature -- in the paramagnetic phase -- we find a sharp line shifted from the Larmor frequency by $\delta\nu$\,=\,$\nu$\,$-\,^{35}\nu_0$\,=\,211(5)\,kHz, whereas three very weak additional lines are resolved when temperature is lowered [Fig.\,\ref{NMR}(a)].
On cooling, all lines with $\delta\nu$\,$>$\,0 shift to higher frequencies, mimicking the behavior of the bulk magnetic susceptibility $\chi_{a^*}(T)$.
For instance, the shift of the dominant line increases and develops a broad maximum $\delta\nu$\,=\,310(5)\,kHz at 50\,K, which is followed by a reduction, $\delta\nu$\,=\,286(5)\,kHz at $T_{N1}$ [Fig.\,\ref{NMR}(b)].
In contrast, the corresponding linewidth does not change significantly down to 50\,K and exhibits a pronounced broadening only below 30\,K, i.e., still far above $T_{N1}$.
This is in line with $\mu$SR results and is reminiscent of the behavior in several frustrated and low-dimensional magnetic systems,\cite{Linarite} where short-range magnetic correlations typically emerge at temperatures that are significantly elevated compared to the magnetic ordering temperature.\cite{Mendels}
In particular, such correlations introduce a distribution of local magnetic fields that effectively broadens the line and potentially influences the line shift.
We note that in some highly frustrated (kagome) systems the broadening of the NMR line with decreasing temperature has been ascribed to a broad distribution of local susceptibilities induced by nonmagnetic defects.\cite{NMRimpurity}
However, in these materials the concentration of defects is high, typically affecting $\sim$5\,\% of the magnetic sites, i.e., orders of magnitude more than in our crystals, where we found no trace of defects/impurities by any experimental technique.
Hence, the defect scenario in FTOC is highly unlikely.
Finally, we stress that short-range correlations were detected also in FTOB  by neutron diffraction as well as by $\mu$SR at $T$\,$<$\,50\,K.\cite{PregeljPD, Zorko}

Below $T_{N1}$ all spectral lines abruptly broaden and develop a box-like shape, indicative of a distribution of local magnetic fields or EFG encountered in the IC systems,\cite{Blinc} as found in FTOB.\cite{PregeljNMR}
The signal at $\delta\nu$\,=\,1\,MHz is now almost twice as broad as the rest of the lines [Fig\,\ref{NMR}(a)].
In addition, on cooling below 12\,K this signal retains the box-like shape, while the rest of the lines develop a triangular form.
Hence it must correspond to a nucleus, which has a different local environment, either due to a different EFG tensor and/or because of a peculiar distribution of local magnetic fields.\cite{PregeljNMR}
At 4\,K the spectrum changes even more and seems to be composed of four broad overlapping lines, centered at the paramagnetic positions.
This must be related to the $T_{N2}$ transition, even though its presence above 7\,T is not clear from the specific heat data.

To clarify which line corresponds to which Cl site, we point out that in FTOB Br$_2$ has much smaller EFG than Br$_1$,\cite{PregeljNMR} implying that the signal with the smallest $|\delta\nu|$ corresponds to Cl$_2$.
We thus ascribe the sharp line at $\delta\nu$\,=\,210\,kHz to the central transition (--1/2\,$\leftrightarrow$\,1/2) of the Cl$_2$ site, while the three weak lines must be the satellite transitions ($\pm$3/2\,$\leftrightarrow$\,$\pm$1/2) of Cl$_2$ and Cl$_1$.
Moreover,  below $T_{N1}$ the line with the shift of $\delta\nu$\,=\,1\,MHz shows distinctly different behavior from the rest, indicating that it corresponds to Cl$_1$, while the two with $\delta\nu$\,=\,$-1$ and 1.5\,MHz correspond to Cl$_2$.
We note that the second satellite as well as the central lines for Cl$_1$ are out of our experimental frequency range.
The room temperature $^{35}$Cl shifts ($\delta\nu$) for Cl$_2$ are approximately 10 times smaller compared to the shifts of $^{79,81}$Br for Br$_2$ in FTOB.
Since $Q_{\text{Br}}$\,$\approx$\,$3Q_{\text{Cl}}$,\cite{NMRtables} $\delta\nu$ scales with $Q^2$, implying that it is dominated by the second order quadrupole shift.\cite{Slichter}
Strikingly, scaling $\delta\nu$ with the magnetic susceptibility reveals that $^{35}$Cl hyperfine shifts for Cl$_2$ are also by an order of magnitude smaller compared to $^{79,81}$Br shifts for Br$_2$ in FTOB.
This, on the other hand, indicates different strengths of the magnetic interactions and probably reflects the difference in the core polarization hyperfine fields of the Br$^-$ and the Cl$^-$ ions.

\subsection{Neutron diffraction experiments}

\begin{figure} [!]
\includegraphics[width=0.50\textwidth]{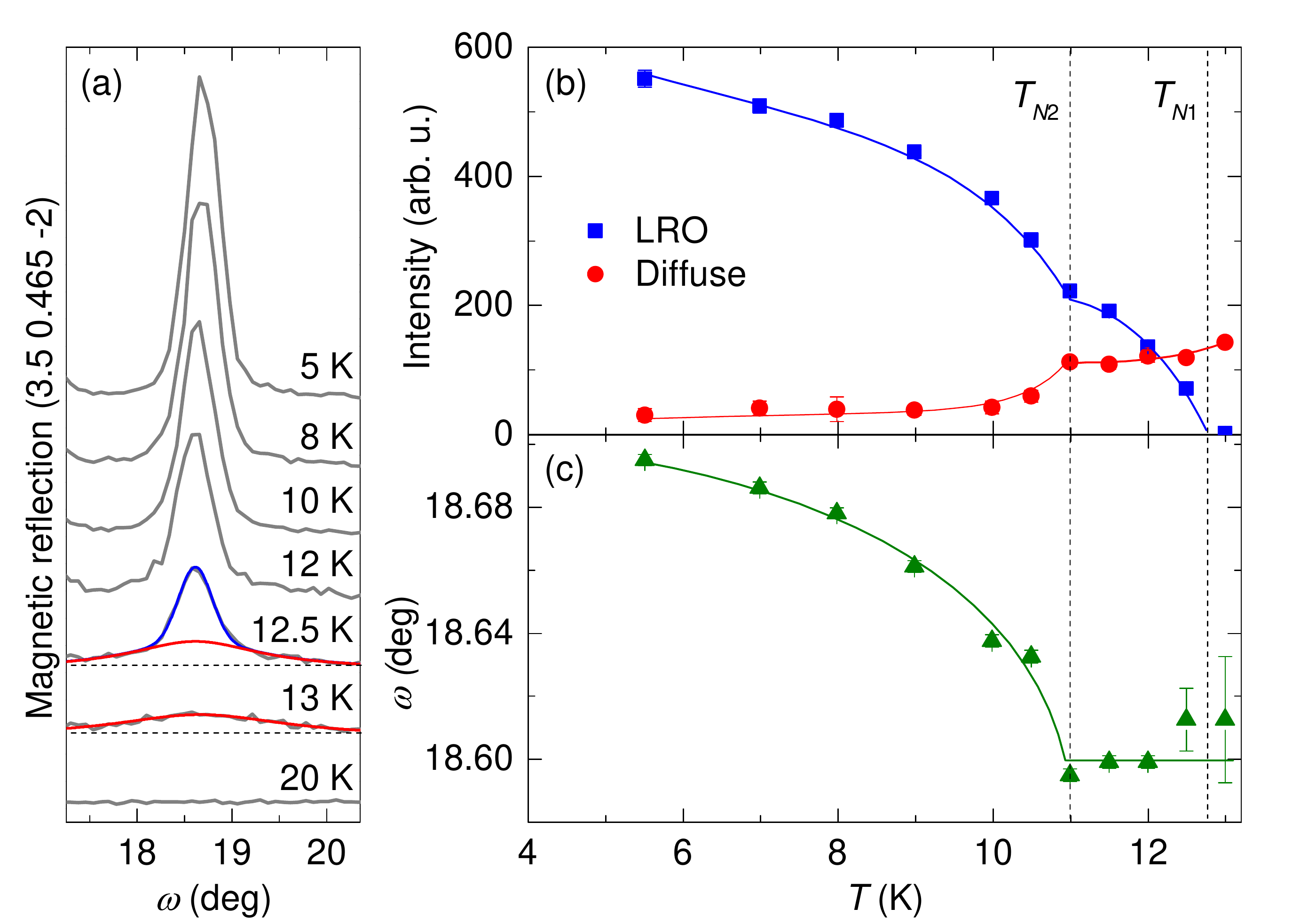}
\caption{(Color online) (a) Temperature evolution of the magnetic reflection (3.5~0.465~-2). The dark (blue) thin solid line is a fit to the pseudo-Voigt function, while the light (red) thin line denotes its Lorentzian part. (b) Derived temperature dependences of the LRO (Gaussian) and diffuse (Lorentzian) magnetic scattering intensity contributions and  (c) the $\omega$ position for this reflection. Solid lines are guides for the eyes.}
\label{ND-par}
\end{figure}

To determine the magnetic ordering in FTOC we conducted neutron diffraction experiments on a single-crystal sample.
Distinct magnetic reflections are found for IC magnetic wave vector {\bf q}$_{\text{IC}}$\,=\,($\frac{1}{2}$\,0.465\,0) at 5\,K, i.e., almost exactly matching the ($\frac{1}{2}$\,0.463\,0) wave vector in FTOB.
To confirm the two magnetic transitions, we first measured the temperature evolution of the strongest magnetic reflection (3.5~0.465~--2) in the $\omega$-scan geometry [Fig.\,\ref{ND-par}(a)].
We point out that in the vicinity of $T_{N1}$ the Gaussian profile, corresponding to the magnetic LRO, is superimposed on top of a broad diffuse Lorentzian contribution, indicative of short-range magnetic correlations.
To derive the temperature dependences of the peak position and its intensity, we thus fitted the data with pseudo-Voigt function, simultaneously accounting for the LRO (Gaussian) and the diffuse (Lorentzian) magnetic scattering contributions.
Our analysis [Fig.\,\ref{ND-par}(b)] clearly reveals the emergence of LRO contribution at $T_{N1}$\,=\,12.5\,K, while the diffuse scattering can be noticed already at 13\,K, i.e., above $T_{N1}$.
The diffuse fraction is suppressed at $T_{N2}$\,=\,11\,K, i.e., at the second magnetic transition, where the intensity of the LRO contribution exhibits an anomaly [Fig.\,\ref{ND-par}(b)] and the magnetic reflection starts to shift with further decreasing temperature [Fig.\,\ref{ND-par}(c)].
Similar behavior was observed in FTOB, where the diffuse scattering below $T_{N1}$ was ascribed to a finite magnetic correlation length along $a^*$,\cite{PregeljPD} while the shifts of the magnetic reflections were identified as a sign of the exchange striction.\cite{PregeljAFMR}
Since both magnetic phases exhibit IC LRO, we address them as HT-IC (between $T_{N1}$ and $T_{N2}$) and LT-IC (below $T_{N2}$) phases, in analogy with FTOB.

\begin{figure} [!]
\includegraphics[width=0.48\textwidth]{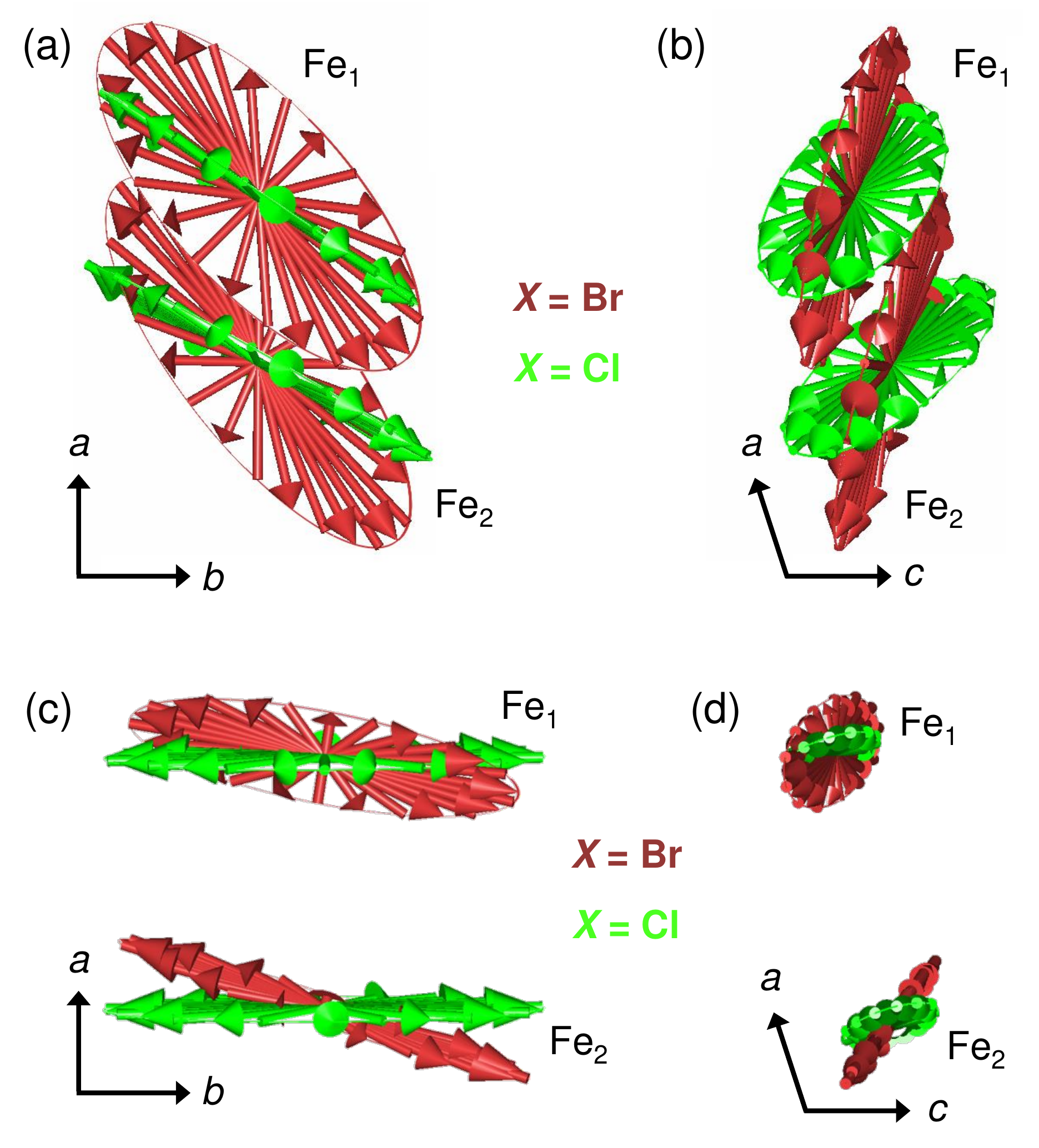}
\caption{(Color online) Magnetic structure models for FeTe$_2$O$_5X$ ($X$\,=\,Br,Cl), compared at Fe$_{11}$ and Fe$_{21}$ sites for the LT-IC phase in (a) $ab$ and (b) $ac$ projections, and for the HT-IC phase in (c) $ab$ and (d) $ac$ projections. For clarity, magnetic unit cell along $b$ is flattened and the sizes of the magnetic moments in the HT-IC phase are magnified by a factor 3.}
\label{structure}
\end{figure}
We next measured intensities of 320 magnetic reflections at 6\,K, which allow for the refinement of the magnetic structure in the LT-IC phase.
As in FTOB, the representation analysis shows that the magnetic wave vector {\bf q}$_{\text{IC}}$\,=\,($\frac{1}{2}$\,0.465\,0) breaks the inversion symmetry, leaving two possible one-dimensional irreducible representations of the little (magnetic) group, which couple magnetic moments at the Fe sites related by a 2$_{1y}$ twofold screw axis.\cite{PregeljPRL09}
In the most general case, the magnetic moment at a particular Fe site is defined as
\begin{equation}
{\bf{S}}_{mn}({\bf{r}}_{i})  =  {\bf{S}}_{0\,mn}^{\text{Re}}\cos({\bf{q}}\cdot{\bf{r}}_{i} - \psi_{mn}) 
 +  {\bf{S}}_{0\,mn}^{\text{Im}}\sin( {\bf{q}}\cdot{\bf{r}}_{i} - \psi_{mn}). 
\end{equation} 
Here, the vector ${\bf{r}}_i$ defines the origin of the $i$-th unit cell, $m$\,=\,1,2 identifies the crystallographically non-equivalent Fe-sites, and $n$=1-4 denotes the four Fe positions within the crystallographic unit cell (for details see Table\,\ref{tab1}).
The complex vector ${\bf{S}}_{0\,mn}$ is determined by its real and imaginary components, ${\bf{S}}_{0\,mn}^{\text{Re}}$ and ${\bf{S}}_{0\,mn}^{\text{Im}}$, which define the amplitude and the orientation of the magnetic moments, i.e., the envelope of the magnetic cycloid/spiral, while $\psi_{mn}$ denotes its phase shift. 
The magnetic wave vector {\bf q} is in units of (2$\pi/a$,\,2$\pi/b$,\,2$\pi/c$) and $\psi_{mn}$ in 2$\pi$.
Since the refinements considering only one irreducible representation did not converge, we took as a starting point the elliptical IC structure model used to describe the LT-IC phase in FTOB.\cite{PregeljPRL12}
To avoid overparameterization of the problem we assume the same complex vector ${\bf{S}}_{0\,mn}$\,$\equiv$\,${\bf{S}}_{0\,m}$ for all crystallographically equivalent Fe sites.
As anticipated, the refinement yields a stable solution, which has strong sinusoidal modulation with $|{\bf S}_{0\,m}^{\text{Im}}|$\,$\sim$\,3$|{\bf S}_{0\,m}^{\text{Re}}|$ and the dominant components ({\bf S}$_{0\,m}^{\text{Im}}$) of the magnetic moments aligned close to the (1~$-1$~0) orientation [Fig.\,\ref{structure}, Table\,\ref{tab1}], as in FTOB.
However, the perpendicular components (spanning the elliptical envelopes) have a leading $c$ component, making the normal of the ellipse almost perpendicular to the one found in FTOB (Fig.\,\ref{structure}), i.e., $\Delta\phi$\,=\,75(4)\,$^\circ$.
To test the reliability of our solution and its deviation from the FTOB magnetic structure, we performed additional refinement, assuming the same ${\bf{S}}_{0\,m}$ as in FTOB and refining only magnetic phases.
The cost of the obtained solution has doubled, which indicates that the normal of the ellipse is indeed rotated compared to FTOB.

\begin{table} [!]
\caption{Components of vectors {\bf S}$_{0\,m}^s$\,=\,($S_{0\,x}^s$,\,$S_{0\,y}^s$,\,$S_{0\,x}^s$) for $s$\,=\,Re,\,Im, defining the elliptical envelops for two independent magnetic atoms (Fe$_1$ and Fe$_2$) for the best magnetic structure model at 6\,K, and eight magnetic phases $\psi_{mn}$ in units of 2$\pi$, i.e., one for each of the magnetic Fe$_{mn}$ atoms in the unit cell ($m$\,=\,1,2, $n$\,=\,1-4). The sites Fe$_{12}$-Fe$_{14}$ are obtained from Fe$_{11}$ [$0.1204(6)$, $0.0008(8)$, $0.9774(7)$] and Fe$_{22}$-Fe$_{24}$ from Fe$_{21}$ [$0.9360(6)$, $0.2953(1)$, $0.8556(6)$] by symmetry elements $i$, $2_{1y}$ and $2_{1y}i$, respectively. The orientation of the moments is given in the $a^*bc$ coordinate system, while $|{\bf S}_{0}|$\,$\approx$\,4\,$\mu_B$.
\label{tab1}}
\begin{ruledtabular}
\begin{tabular}{ccccc}
     $s$ = Re, Im   & Fe$_1^{\text{Re}}$ & Fe$_1^{\text{Im}}$ & Fe$_2^{\text{Re}}$ & Fe$_2^{\text{Im}}$   \\
\hline
$S_{0\,x}^s$/$|{\bf S}_{0\,m}^s|$        &-0.31   & 0.49   &  -0.40 &    0.39       \\
$S_{0\,y}^s$/$|{\bf S}_{0\,m}^s|$        & 0.18   &-0.81   &   0.26 &   -0.81       \\
$S_{0\,z}^s$/$|{\bf S}_{0\,m}^s|$        & 0.93   & 0.31   &   0.88 &    0.42       \\ 
\hline
$|{\bf S}_{0\,m}^s|/|{\bf S}_{0}|$  & 0.34   & 0.92   &  0.27  &  1.00  \\
\hline
\hline
 $m$ &  $\psi_{m1}$ & $\psi_{m2}$ & $\psi_{m3}$ & $\psi_{m4}$\\
\hline
 1     &  0.00            &  0.12          &    0.23         &     0.88 \\
 2     &  0.07            &  0.95          &    0.25         &     0.13  \\
\end{tabular}
\end{ruledtabular}
\end{table}

\begin{table} [!]
\caption{Parameters for the structure model for the HT-IC phase for the data measured at 11.2\,K with $|{\bf S}_{0}|$\,$\approx$\,1.2\,$\mu_B$. For details see caption of Table\,\ref{tab1}.
\label{tab2}}
\begin{ruledtabular}
\begin{tabular}{ccccc}
     $s$ = Re, Im   & Fe$_1^{\text{Re}}$ & Fe$_1^{\text{Im}}$ & Fe$_2^{\text{Re}}$ & Fe$_2^{\text{Im}}$   \\
\hline
$S_{0\,x}^s$/$|{\bf S}_{0\,m}^s|$        &-0.05   &-0.01  &-0.05   &-0.01       \\
$S_{0\,y}^s$/$|{\bf S}_{0\,m}^s|$        & 0.05   & 1.00  & 0.05   & 1.00      \\
$S_{0\,z}^s$/$|{\bf S}_{0\,m}^s|$        &-1.00   &-0.03  &-1.00   &-0.03        \\ 
\hline
$|{\bf S}_{0\,m}^s|/|{\bf S}_{0}|$  & 0.21   & 1.00   &  0.21  &  1.00  \\
\hline
\hline
 $m$ &  $\psi_{m1}$ & $\psi_{m2}$ & $\psi_{m3}$ & $\psi_{m4}$\\
\hline
 1     &  0.00            &  0.12          &    0.34         &     0.45 \\
 2     &  0.23            &  0.10          &    0.42         &     0.30  \\
\end{tabular}
\end{ruledtabular}
\end{table}

Next, we collected 140 magnetic reflections at 11.2\,K, i.e., in the HT-IC phase.
The best solution is again almost completely sinusoidal model, with magnetic moments pointing along the $b$ axis (Table\,\ref{tab2}).
The shorter component of the elliptical envelope points  here exactly along the $c$ axis, keeping the normal of the ellipse perpendicular to the one in FTOB also in the HT-IC phase.\cite{PregeljNMR}
We note that due to the limited amount of data and weak magnetic reflections, we assumed further simplification $|{\bf S}_{0m}|$\,$\equiv$\,$|{\bf S}_{0}|$.
Once more, the refinements with a single irreducible representation do not converge.
Finally, in contrast to FTOB no systematic changes of magnetic phases $\psi_{mn}$ were found, implying that electric polarization in FTOC might occur already in the HT-IC phase.

\subsection{Dielectric response}

\begin{figure} [!]
\includegraphics[width=0.50\textwidth]{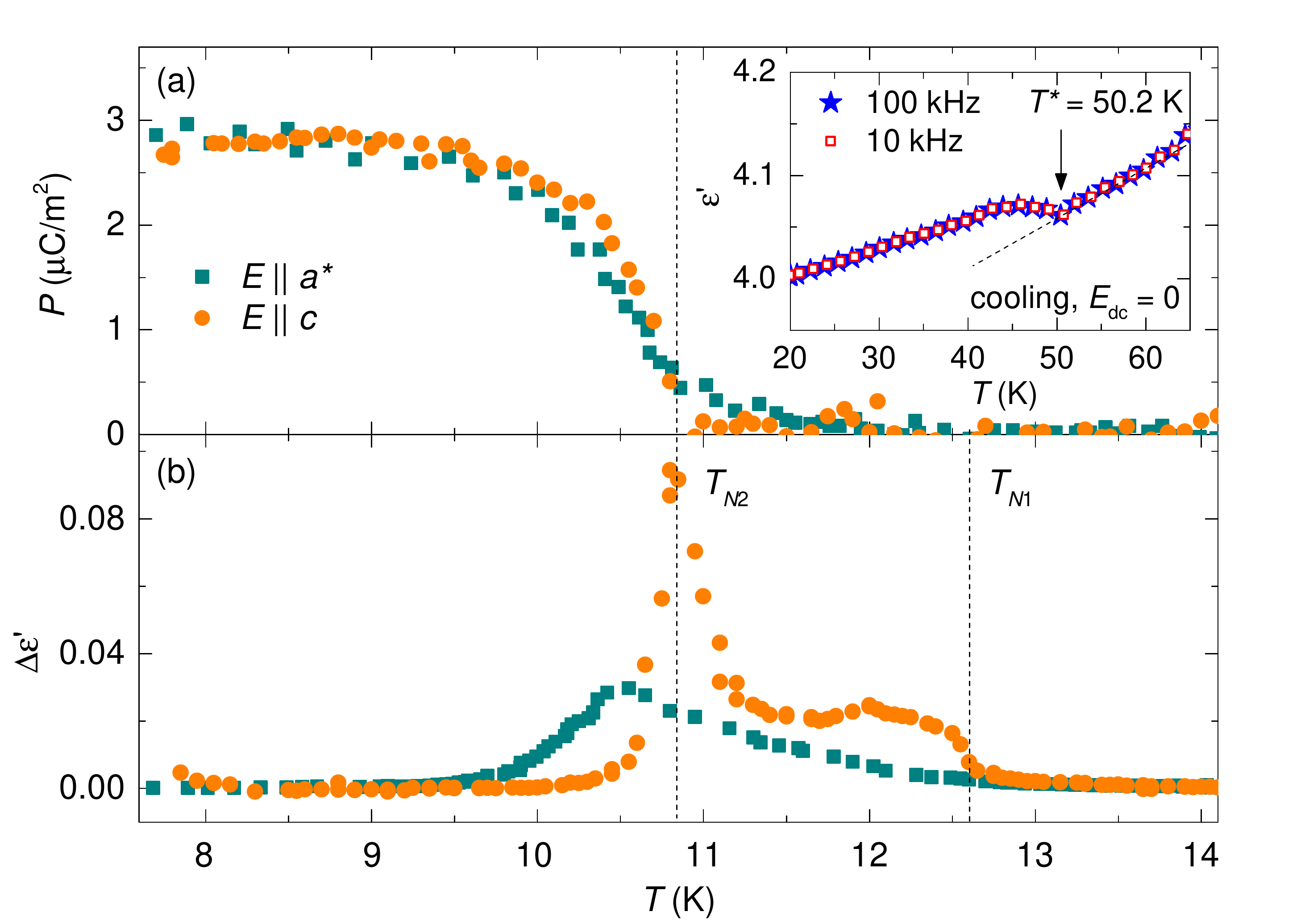}
\caption{(Color online) Temperature dependences of (a) the electric polarization for electric field $E$\,=\,10\,kV/cm applied along $a^*$ and $c$ axes and changes of (b) the corresponding dielectric constants measured in ac field oscillating with frequency $\nu_{\text{ac}}$\,=\,10\,kHz. Inset: an anomaly in the $\epsilon'(T)$ for $E_{ac}||c$ measured at two different $\nu_{\text{ac}}$ demonstrating its frequency independence.}
\label{pol}
\end{figure}
To test the ferroelectric response of FTOC, we measured temperature dependences of the dielectric constant $\epsilon'$ and the electric polarization, $P$, on the single crystal samples in electric fields, $E$, applied along each of the three crystallographic axes ($a^*$,\,$b$,\,$c$).
On cooling from room temperature, $\epsilon_c'(T)$ ($E_{\text{ac}}||c$) first displays a clear anomaly at $T^*$\,=\,50.2(1)\,K [Inset in Fig.\,\ref{pol}(a)], which is close to the maximum in the magnetic susceptibility\cite{Becker} that corresponds to the onset of short-range magnetic correlations, also denoted by the broadening of the NMR line.
No further dielectric anomalies were found until $T_{N1}$, where $\Delta\epsilon_c'(T)$ exhibits a clear step, which is followed by a pronounced $\lambda$-type anomaly at $T_{N2}$ [Fig.\,\ref{pol}b].
In contrast, $\Delta\epsilon_{a^*}'(T)$ reveals no anomaly at $T_{N1}$, but has a broad ($\sim$3\,K) $\lambda$-type anomaly centered at $\sim$($T_{N2}-0.5$\,K).

In spite of the fact that $\Delta\epsilon_c'(T)$ exhibits anomaly already at $T_{N1}$, finite net electric polarization for $E||c$ emerges only below $T_{N2}$ (Fig.\,\ref{pol}).
The response for $E||a^*$, however, is not so clear, as the corresponding $P$ builds up already 0.5-1\,K above $T_{N2}$.
Nevertheless, below $T_{N2}$, $P(T)$ for $E||a^*$ and $E||c$ exhibits a typical order-parameter-like behavior, reaching a saturation at $\sim$8\,K.
The maximum values for both field orientations are the same $P$\,=\,2.8(1)\,$\mu$C/m$^2$ and amount to roughly one third of the value acquired in FTOB.
We note that for $E||b$ $P$\,=\,0 was found in the entire temperature region.

\subsection{X-ray diffraction}

\begin{figure} [!]
\includegraphics[width=0.48\textwidth]{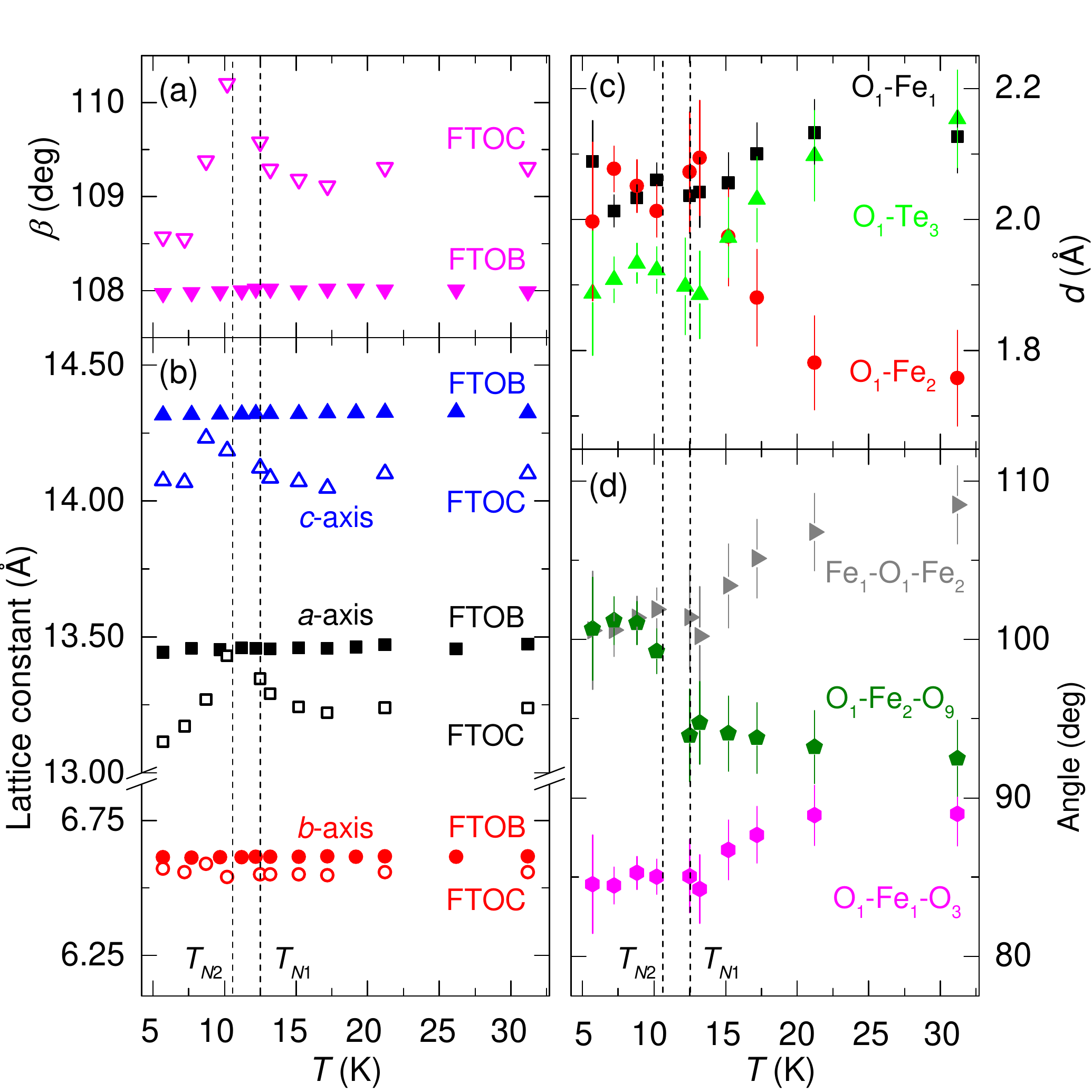}
\caption{(Color online) Temperature dependences of (a,b) the lattice parameters for the two isostructural FeTe$_2$O$_5X$ ($X$\,=\,Br, Cl) compounds. The error bars are smaller than the size of the symbols. Additional temperature dependences of (c) the interatomic distances and (d) the angles, exhibiting most pronounced changes in the vicinity of $T_{N1}$ for the FeTe$_2$O$_5$Cl (FTOC).}
\label{x-ray-all}
\end{figure}

Finally, we inspected the response of the crystal lattice by performing synchrotron x-ray diffraction on a single-crystal sample.
To refine the crystal structure successfully, high-temperature (paramagnetic) crystal symmetry (space group $P$2$_1$/$c$) was considered at all temperatures, hence all the deduced crystallographic changes are centrosymmetric and provide only indirect evidences of non-centrosymmetric ferroelectric distortions.
The changes of the lattice parameters at $T_{Ni}$ are very pronounced [Fig.\,\ref{x-ray-all}(a),(b)], in contrast to FTOB, where these effects were minute.
The most noticeable anomalies at $T_{N2}$ are found for $\beta$, $a$ and $c$, which are all related to the interlayer spacing.
On the other hand,  the temperature-independent $b$ appears to be insensitive to the magnetic ordering.
We stress that the crystal layers in this compound are formed due to Te$^{4+}$ lone-pair cations.\cite{Becker, ChemSci}  
Consequently, all changes in FTOC that are associated with the interlayer distance most likely reflect the shift of the Te$^{4+}$ lone-pair electrons, which is also the most probable origin of the electric polarization.\cite{PregeljPRL09,PregeljAFMR}

\begin{figure} [!]
\includegraphics[width=0.48\textwidth]{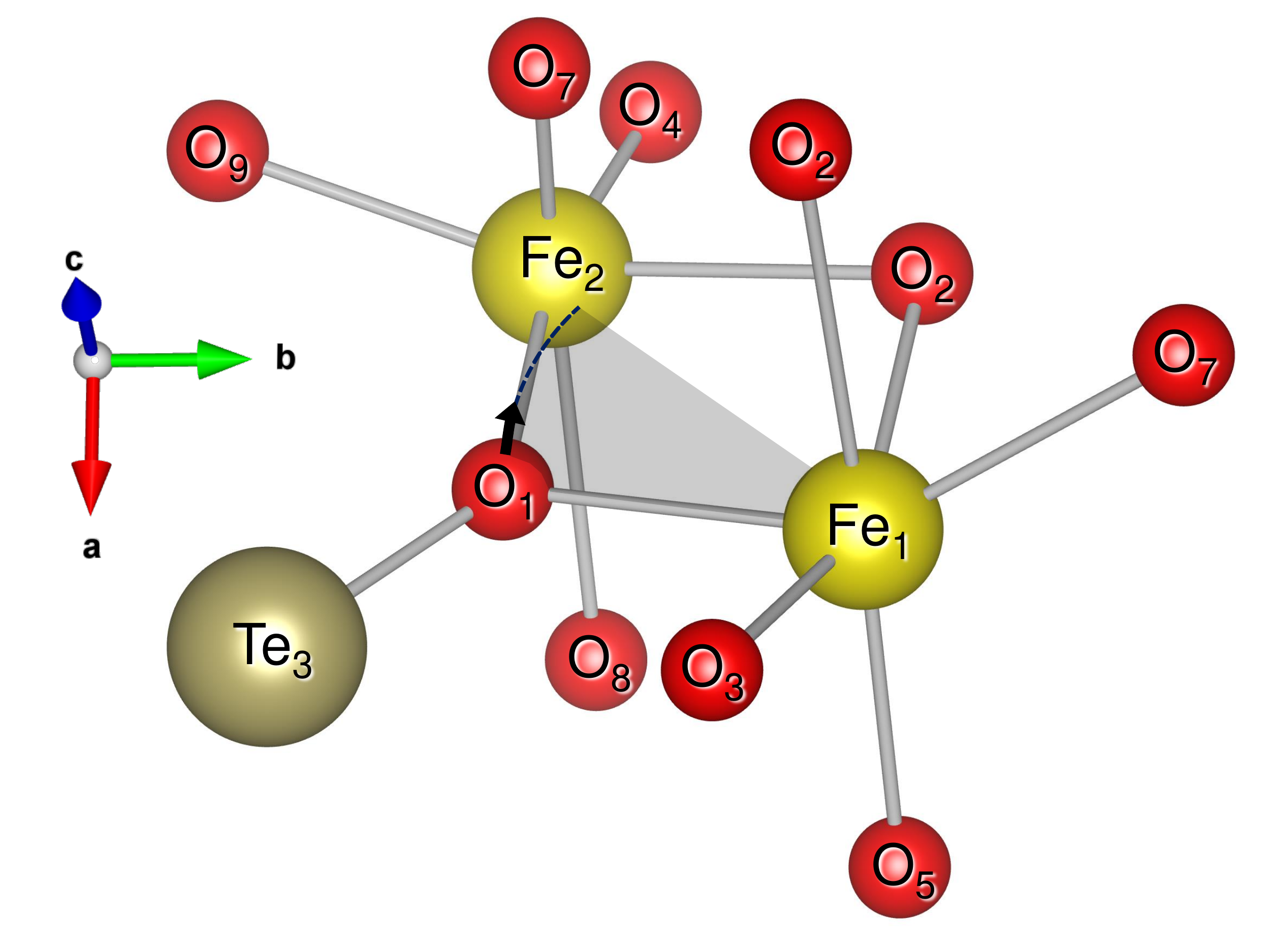}
\caption{(Color online) Local O$_1$ surrounding and assumed direction (marked by black arrow) of the O$_1$ shift on heating through the $T_{N2}$ transition.}
\label{O1-struc}
\end{figure}

Our refinement further enables the extraction of the interatomic distances and thus allows us to identify the atoms involved in the ME effect.
We find that all significant changes close to $T_{N2}$ are associated with the O$_1$ atom.
In Fig.\,\ref{x-ray-all}(c) we show temperature dependences of the distances between the O$_1$ and the neighboring Fe and Te atoms that it binds (Fig.\,\ref{O1-struc}).
All other changes of the interatomic distances within the crystal layer are much smaller (not shown), as they are below our experimental sensitivity.
The fact that the O$_1$-Fe$_2$ and the O$_1$-Te$_3$ distances exhibit significantly larger changes than the O$_1$-Fe$_1$ one implies that the O$_1$ atom is actually rotating around the Fe$_1$ one (Fig.\,\ref{O1-struc}).
This reflects also in the angles between the O$_1$ and the neighboring O and Fe atoms shown in Fig.\,\ref{x-ray-all}(d).
In particular, the changes of the O$_1$-Fe$_2$-O$_9$ and the Fe$_1$-O$_1$-Fe$_2$ angles are larger than that of the O$_1$-Fe$_1$-O$_3$ one, which further implies that the axis of the O$_1$ rotation is close to the Fe$_1$-O$_3$ direction.
Since changes of other interatomic angles appear to be much smaller (not shown), the observed rotation most likely applies to the O$_1$ atom alone and does not involve the whole Fe$_1$-O octahedron.
We thus suggest that this rotation is a direct response to the lone-pair redistribution on the Te$_3$ ion.

\section{Discussion}

At first sight, the only difference between the FTOC system investigated here and its isostructural relative FTOB is the species of the halogen ion.
Since halogen ions are located in-between the FeTeO layers, they mostly affect the interlayer distance.
The dominant intralayer magnetic exchange interactions in the two systems are thus expected to be comparable and should lead to similar magnetic orders.
Nevertheless, the difference in the interlayer distance may significantly influence the ME coupling, as Te$^{4+}$ lone-pair electrons, which are involved in the Fe-O-Te-O-Fe magnetic exchange pathway and most probably carry the electric polarization,\cite{PregeljAFMR} protrude into the interlayer space.\cite{Becker}
Following, we summarize our main findings and discuss how they comply with the above assumptions.

First, we stress that we find a finite ferroelectric polarization and clear evidences of the ME effect, in contrast to the prior far-infrared study\cite{Pfuner} that could not detect any link between electrodynamic response and magnetic ordering.
The observed dielectric response may be explained in the following way.
At $T_{N1}$, small electric dipoles develop along the $c$ axis, inducing a concurrent step in $\epsilon_c'(T)$ [Fig.\,\ref{pol}(b)].
However, since $P(T)$ for $E||c$ is flat down to $T_{N2}$ [Fig.\,\ref{pol}(a)], these dipoles are most likely ordered antiparallelly, leading to the antiferroelectric polarization $P_c^{AF}$.
When $E||a^*$ is applied, $P_c^{AF}$ bends towards $a^*$, resembling a finite net polarization already in the HT-IC phase [Fig.\,\ref{pol}(a)].
At $T_{N2}$, a "metaelectric" transition occurs, where antiferroelectrically ordered dipoles flop and form a ferroelectric arrangement pointing along $c$, $P_c^{F}$, reflecting as a rapid increase of $P$ for $E||c$ and a sharp anomaly in $\epsilon_c'(T)$.
On further cooling, $P(T)$ for $E||{a^*}$ almost exactly follows the response for $E||c$, suggesting that spontaneously induced electric polarization $P_c^{F}$ rotates towards $a^*$ when $E||{a^*}$  is applied.

To address the magnetic behavior of the FeTe$_2$O$_5X$ system, we first highlight both characteristic temperatures, i.e., the Curie-Weiss ($\theta$) and the N\'{e}el ($T_{N1}$) temperature, which are in FTOC $\sim$25\,\% higher than in FTOB.\cite{Becker}
In the case of a non-frustrated layered system, $T_{N}$ is related to both, intralayer as well as interlayer, exchange interactions,\cite{Pokrovsky1990, Yasuda2005} whereas $\theta$ is determined solely by dominant intralayer interactions.
On the contrary, in frustrated systems, $T_{N}$ is determined by main competing interactions\cite{Mendels} and should thus scale proportionally to $\theta$.
The fact that the substitution of the halogen ion, i.e., the change of the interlayer distance, induces similar relative changes of both characteristic temperatures suggests that in FTOC/FTOB frustration dominates and overcomes the layer-type effects.
This is in agreement with antiferromagnetic resonance and density functional theory study, revealing a quasi one-dimensional character of the system.\cite{PregeljAFMR}
Nonetheless, the reduction of the interlayer distance may still cause higher-order effects and thus influence the magnetic anisotropy and/or the ME coupling.

Indeed, the two systems exhibit distinctly different magnetic anisotropies, as their flat elliptical envelopes are almost perpendicular to each other in both LRO phases [Fig.\,\ref{structure} and Tables\,\ref{tab1},\,\ref{tab2}].\cite{PregeljNMR}
This further reflects in the magnetic-field dependence of $T_{N2}$, which is in both systems reduced when field is applied along $a^*$ (Fig.\,\ref{cp}),\cite{PregeljPD, PregeljNMR} but the effect is more significant in FTOC.
The reason for the observed response is that the magnetic field prefers antiferromagnetically ordered magnetic moments (forming a cycloid) to point perpendicular to it, making the LT-IC ordering with the dominant orientation of the magnetic moments along (1\,$-1$\,0) unfavorable.\cite{PregeljPD, Nagayima}
The larger decrease of $T_{N2}$ in FTOC, on the other hand, complies with the orientation of the HT-IC elliptical envelope that is in FTOC perpendicular to $a^*$, whereas in FTOB the normal of the ellipse points towards $c$. 
Such a pronounced anisotropic behavior is rather surprising, as the magnetic Fe$^{3+}$ ion in the high-spin $S$\,=\,5/2 state has an almost isotropic orbital ground state ($L$\,=\,0), for which crystal-field anisotropy is expected to be small.
Distinctly different anisotropies of FTOC and FTOB, therefore, imply that the two systems are highly sensitive to the weak Fe$^{3+}$ magnetic anisotropy.
This is most likely due to the geometric frustration, responsible for the complex magnetic ground state, which is probably very close to being degenerated and thus also very sensitive to small perturbations.

Due to the ME coupling, the magnetic anisotropies are most likely responsible also for the discrepancy between the dielectric responses of the two systems.
In particular, the measured electric polarization in FTOC is in the LT-IC phase equal for $E||a^*$ and $E||c$, whereas in FTOB the response for $E||a^*$ is almost an order of magnitude weaker than for $E||c$.
This may indicate that in FTOC the spontaneously induced net electric polarization $P_c^F$ can be easily rotated by the external electric field, while in FTOB it is almost completely insensitive to it.
Similarly, finite $P$ for $E||a^*$ above $T_{N2}$ may result from a potentially existing antiferroelectric component $P_c^{AF}$, which is bend towards $a^*$ due to the applied electric field.
Our results thus hint that the electric polarization in FTOC is significantly less rigid than in FTOB and can be easily rotated within the $ac$ plane by the external electric field, potentially affecting also the magnetic ordering.
This further complies with the magnetic-field dependence of $T_{N2}$, which indicates that anisotropies in FTOC are weaker than in FTOB.

Finally, we look at the ME coupling at the microscopic scale.
In contrast to FTOB, crystal-lattice changes at $T_{N2}$ are in FTOC much more distinct and clearly indicate the shift of the O$_1$ atom.
We stress here that O$_1$ is bridging $J_2$ Fe-O-Fe as well as the $J_5$ Fe-O-Te-O-Fe exchange interactions,\cite{PregeljAFMR} from which $J_2$ was identified as the strongest interaction in FTOB, while $J_5$ includes Te$_3$ -- the only Te$^{4+}$ ion showing a noticeable shift in FTOB.\cite{PregeljPRL09}
Moreover, $J_5$ was found to have very strong influence on the electric polarization.\cite{Chakraborty}
Since O$_1$ is involved in a strong exchange interaction as well as it is attached to the Te$^{4+}$ ion, which has easily polarizable lone-pair electrons, the crystallographic changes at $T_{N2}$ (Fig.\,\ref{x-ray-all}) suggest that the O$_1$ atom is the source of the ME effect in FTOC/FTOB.

\section{Conclusions}

A broad set of experimental methods was used to investigate magnetic and dielectric response of the FTOC system.
We find that the system is multiferroic and is in many aspects very similar to its isostructural relative FTOB.
In particular, it exhibits two IC magnetically ordered states with very similar elliptical cycloidal magnetic orders and nearly the same dominant orientations as in FTOB.
However, the cycloidal planes are substantially tilted, i.e., the one in FTOC is almost perpendicular to the one in FTOB.
Since the only clear crystallographic difference between the two systems involves the interlayer distance, the observed response is ascribed to the geometric frustration.
The latter most likely amplifies the effects of otherwise subtle structural differences, which are probably responsible for the variation of the magnetic anisotropies as well as for the different responses of the electric polarization to the applied electric field.
On the microscopic level, however, the ME mechanism seems to involve O$_1$, which bridges the main exchange interaction as well as it is attached to the Te$^{4+}$ ion, which has easily polarizable lone-pair electrons that perturb into the interlayer spacing.
Hence, lone-pair electrons may be affected by distortions of the exchange bridges through magnetostriction process as well as by a variation of the interlayer distance, which can thus both influence the response of the electric polarization. 
Finally, the observed persistent spin dynamics yet again confirms that this phenomenon is intrinsic to amplitude-modulated magnetic structures.

\acknowledgments
We acknowledge the financial support of the Slovenian Research Agency (project J1-2118 and Z1-5443) and the Swiss National Science Foundation (project No.~200021-129899).
$\mu$SR measurements at the ISIS facility have been supported by the European Commission under the 7th Framework Program through the ``Research Infrastructures'' action of the ``Capacities''  Program, NMI3-II Grant No. 283883. Contract No. 283883-NMI3-II. The neutron diffraction work was performed at SINQ, Paul Scherrer Institute, Villigen, Switzerland.

\end{document}